\documentclass[twocolumn]{aastex62}

\usepackage[USenglish]{isodate}
\usepackage{wasysym}
\usepackage[english,greek]{babel}

\usepackage{comment}

\newcommand\gaia{\textit{Gaia}}

\def\bmr{$(G_{\rm BP}-G_{\rm RP})$}
\def\mg{$M_G$}

\received{May 29, 2018}
\accepted{April 16, 2019}
\submitjournal{ApJ}

\shorttitle{Magnetic transitions in late-type dwarfs}
\shortauthors{Lanzafame et al.}

\begin{document}
\selectlanguage{English}

\title{Evidence of New Magnetic Transitions in Late-Type Dwarfs from \textit{Gaia} DR2}

\correspondingauthor{Alessandro C. Lanzafame}
\email{a.lanzafame@unict.it}

\author[0000-0002-2697-3607]{Alessandro C. Lanzafame}
\affil{Universit\`a di Catania, Dipartimento di Fisica e Astronomia, Italy }
\affiliation{INAF-Osservatorio Astrofisico di Catania, Italy}

\author{Elisa Distefano}
\affiliation{INAF-Osservatorio Astrofisico di Catania, Italy}
\nocollaboration

\author{Sydney A. Barnes}
\affiliation{Leibniz Institute for Astrophysics (AIP), Potsdam, Germany}
\nocollaboration

\author{Federico Spada}
\affiliation{Max Planck Institute for Solar System Research, G\"ottingen, Germany}
\nocollaboration



\begin{abstract}
The second \gaia\ data release contains the identification of 147\,535 low-mass ($\apprle 1.4 M_{\odot}$) rotational modulation variable candidates on (or close to) the main sequence, together with their rotation period and modulation amplitude.
The richness, the period and amplitude range, and the photometric precision of this sample make it possible to unveil, for the first time, signatures of different surface inhomogeneity regimes in the amplitude-period density diagram.
The modulation amplitude distribution shows a clear bimodality, with an evident gap at periods $P \apprle 2$\,d.
The low amplitude branch, in turn, shows a period bimodality with a main clustering at periods $P \approx$ 5 - 10\,d and a 
secondary clustering of ultra-fast rotators at $P \apprle 0.5$\,d.
The amplitude-period multimodality is correlated with the position in the period-absolute magnitude (or period-color) diagram, with the low- and high-amplitude stars occupying different preferential locations.
Here we argue that such a multimodality represents a further evidence of the existence of different regimes of surface inhomogeneities in young and middle-age low-mass stars and we lay out possible scenarios for their evolution, which manifestly include rapid transitions from one regime to another.
In particular, the data indicate that stars spinning up close to break-up velocity undergo a very rapid change in their surface inhomogeneities configuration, which is revealed here for the first time.
The multimodality can be exploited to identify field stars of age $\sim$ 100 -- 600 Myr belonging to the slow-rotator low-amplitude sequence, for which age can be estimated from the rotation period via gyrochronology relationships.  
\end{abstract}



\section{Introduction} \label{sec:intro}

In the second half of last century, the seminal work of \citet{1958ApJ...128..664P,Weber_Davis:1967,1972ApJ...171..565S,Kawaler:1988,Kawaler:1989} opened a new frontier in deriving stellar age from rotation, which represents an appealing alternative to traditional dating methods that use stellar properties that either change little as the stars age or are hard to measure \citep[e.g.][and references therein]{Meibom_etal:2015}.
Observations of young open clusters, however, revealed a wide dispersion of rotation rates, mainly attributed to initial conditions and different disk interaction lifetime during the stellar contraction towards the main sequence, that seemed to prevent the possibility of obtaining sufficient accuracy from dating methods based on rotation \cite[e.g.][]{1993AJ....106..372E}.
The power-law dependence of the angular momentum loss on the rotational velocity, i.e.  the \citet{Kawaler:1988} $\dot{J} \propto \Omega^3$ wind-braking law, where $J$ is the stellar angular momentum and $\Omega$ the angular velocity, ensures a decrease of the rotation dispersion in time, but this appears insufficient to explain the rotational evolution seen in open clusters younger than the Hyades (600\,Myr).

The behaviour of young rapidly rotating stars suggested, in fact, a weaker dependence of $\dot{J}$ on $\Omega$ \citep{1987ApJ...318..337S}.
This prompted \citet{1995ApJ...441..865C,1995ApJ...441..876C} to introduce a modification of the \citet{Kawaler:1988} wind-braking law of the type $\dot{J} \propto \Omega_{\mathrm{sat}}^2\Omega$ for $\Omega \ge \Omega_{\mathrm{sat}}$, with $\Omega_{\mathrm{sat}}$ introducing a saturation  level into the angular momentum loss law.
Regarding the value of $\Omega_{\mathrm{sat}}$, \citet{Krishnamurthi_etal:1997} proposed the scaling law
\begin{equation}
\label{eq:Krishna_scaling}
(\Omega_{\mathrm{sat}} / \Omega_{\mathrm{sat}\odot}) = (\tau / \tau_{\odot}) \,,
\end{equation} 
with 
\begin{equation}
\label{eq:Krishna_threshold}
\Omega_{\mathrm{sat}\odot} = 10\, \Omega_{\odot}
\end{equation} 
and $\tau$ the convective turnover timescale.

From another perspective, \citet{Barnes:2003} identified a sequence of slowly rotating stars in the color-period diagram of young open clusters that becomes increasingly prominent with increasing cluster's age. 
Stars in this sequence follow approximately the \citet{1972ApJ...171..565S} $P \propto \sqrt{t}$ law at fixed mass, with $t$ the stellar age, strictly related to the \citet{Kawaler:1988} $\dot{J} \propto \Omega^3$ wind-breaking law, with modifications strongly dependent on mass due to the transfer of angular momentum from the stellar core to the envelope in the first $\approx1$\,Gyr  \citep[see, e.g.][]{1993ApJ...409..624S,2011MNRAS.416..447S,2015A&A...584A..30L}.  
The strong $\dot{J}$ dependence on $\Omega$ on the slow-rotator sequence implies that this sequence correspond to the unsaturated regime in the \citet{1995ApJ...441..865C,1995ApJ...441..876C} description.
It also implies that, as the stars age, their rotation periods converge to a unique surface $P=P(t,M)$ in $(P, t, M)$ space, where $M$ the stellar mass, reducing progressively the initial scatter around this surface \citep[see e.g.][]{Meibom_etal:2015}.

Analysing the observed $P=P(t,M)$ relationship in open clusters, \citet{2015A&A...584A..30L} put constraints on the mass dependence of the wind breaking law in the unsaturated regime.
They found that none of the {\it ab initio} theoretical model proposed so far can be fitted satisfactorily to the data.
Only two semi-empirical models amongst those considered were found to fit the data satisfactorily: one derived from the work of \citet{Barnes:2010} and \citet{Barnes_Kim:2010}, 
\begin{equation}
\label{eq:LS15}
\dot{J} \propto I \tau \Omega^3 \,;
\end{equation}
the other proposed by \citep{Matt_etal:2015}, 
\begin{equation}
\label{eq:Matt15}
\dot{J} \propto 
\left\{
\begin{array}{cc}
M^{0.5} R^{3.1} \tau^2 \Omega^3  & \mathrm{for} \; Ro > Ro_{\odot} / \chi 
\\
M^{0.5} R^{3.1} \tau \Omega_{\mathrm sat} \Omega  & \mathrm{for} \; Ro \le  Ro_{\odot} / \chi 
\end{array}
\right. 
\end{equation} 
with $\chi=10$, derived from a physically motivated scaling for the dependence on Rossby number and an empirical scaling with stellar mass and radius.

The open cluster age at which the clustering on the slow-rotator sequence starts to be noticeable is still uncertain.
This sequence is particularly evident starting from the age of the Pleiades, but there is evidence of a fairly defined sequence as early as the age of the $\beta$ Pictoris association \citep[25\,Myr,][]{2017A&A...607A...3M}.

The duration of the early stellar evolution that is affected by the saturation regime depends on the assumptions made on the saturation threshold. 
When compared with rotational evolution track and colour-period diagrams, Eq.\,(\ref{eq:Krishna_scaling}) and (\ref{eq:Krishna_threshold}) imply that the star may leave the saturation regime before or significantly after its settling on the slow-rotator sequence. 
Stars of approximately solar mass will leave the saturation regime before settling on the slow-rotator sequence, but stars of lower mass will stay in the saturation regime for a significant part of their earlier evolution on the slow-rotator sequence \citep{2015A&A...584A..30L}.
This is obviously inconsistent with the evidence that stars in the slow-rotator sequence are in the unsaturated regime and calls for a more in-depth description of the saturated regime and the transition to the unsaturated regime.
In this regards, it is worth noticing that the \citet{2015A&A...584A..30L} wind-braking law, Eq.\,(\ref{eq:LS15}), applies to all stars in the slow-rotator sequence, while the \citet{Matt_etal:2015} wind-braking law, Eq.\,(\ref{eq:Matt15}), relies on a saturation threshold for the Rossby number which, essentially, is very similar to the \citet{Krishnamurthi_etal:1997} one.

\citet{Barnes:2003} postulated that different types of dynamo operate in fast- and slow-rotators in open clusters younger than the Hyades, with a convective dynamo characterising the fast rotators (C-sequence) and an interface (tachocline) dynamo characterising the slow rotators (I-sequence).
In such a scenario, stars in the C-sequence would switch to the I-sequence at $\approx $100-300\,Myr, depending on mass, converging quite rapidly to the slow-rotator sequence (gap phase).
A closer reproduction of the fast- and slow-rotator distribution in open cluster has been obtained by \citet{2014ApJ...789..101B} by assuming a metastable dynamo that produces stochastic transitions from fast- to slow-regimes.

The relevance of these considerations for very low-mass stars that remain fully convective in the MS is still rather speculative. 
\citet{2015A&A...584A..30L} could not apply their analysis to stars of mass lower than 0.6 $M_{\odot}$ because of lack of period data for the older clusters in the sample.
The information conveyed by Gaia DR2 for these stars is also still insufficient.
The work of, e.g., \cite{2017ApJ...834...85N} indicates that a slow-rotator sequence of H$\alpha$ inactive stars can also be identified for M dwarfs.
A $L_{H\alpha}/L_{\mathrm bol}$ saturation threshold is also found at $Ro=0.2$.
However, the impossibility of tracing the rotational evolution of very low-mass stars over a sufficiently long age range with current data prevents to draw similar considerations regarding the wind braking law of these stars.

In this paper we argue that the period and amplitude of the rotational modulation variable candidates contained in the second \gaia\ data release \citep[DR2][]{2018A&A...616A...1G,2018A&A...616A..16L} unveil a global picture of the modifications of the stellar surface inhomogeneities  that provides further evidence of rapid transitions amongst different surface inhomogeneities regimes in the first 600\,Myr of the evolution of low-mass stars developing a radiative core in the main sequence.

The paper is organised as follows: in Sect.\,\ref{sec:data} we give a description of the \gaia\ DR2 data used in this work and a brief summary of the processing and analysis carried out in the context of the DPAC activities on rotational modulation variables; in Sect.\,\ref{sec:clustering} we discuss more in-depth with respect to \cite{2018A&A...616A..16L} the clustering in the amplitude-period diagrams unveiled by \gaia\ for the first time, comparing them with well studied stars, open clusters, and Kepler data; in Sect.\,\ref{sec:discussion} we propose a new scenario for the stellar magneto-rotational evolution from the T\,Tau phase to the late main sequence spurred by the \gaia\ evidence; in Sect.\,\ref{sec:conclusions} we draw our conclusions.

\section{Data} \label{sec:data}

\begin{figure*}[t!]
\gridline{\leftfig{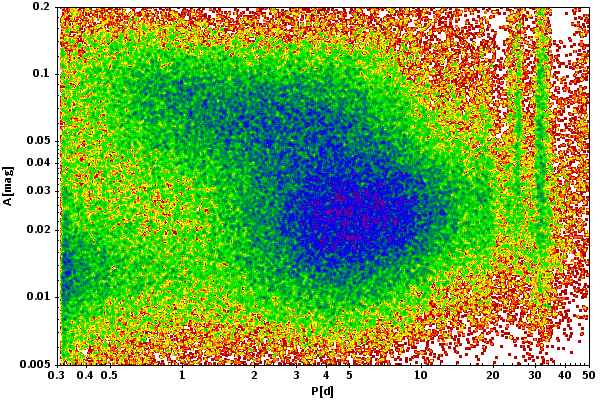}{0.46\textwidth}{(a)}
          \rightfig{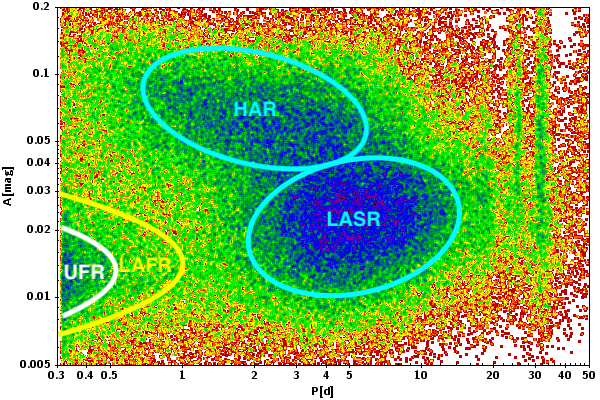}{0.46\textwidth}{(b)}}
\gridline{\leftfig{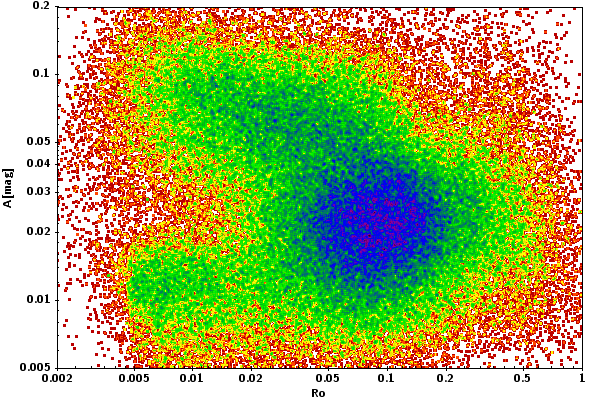}{0.46\textwidth}{(c)}
          \rightfig{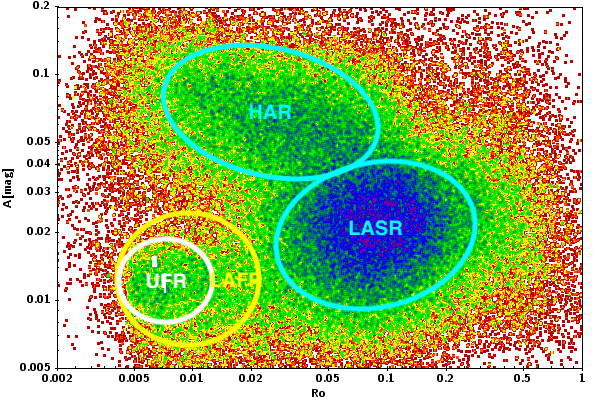}{0.46\textwidth}{(d)}
          }
\caption{Panel (a): amplitude-period density diagram of all the \gaia\ rotational modulation candidates included in DR2. Panel (b): same as panel (a) with the major clustering annotated (HAR high-amplitude rotator; LASR low-amplitude slow-rotator; LAFR low-amplitude fast-rotator; UFR ultra-fast rotator). The UFR group is considered as a sub-group of LAFR. Panels (c): amplitude-$Ro$ (Rossby number) for the same data of panel (a). Panel (d) annotated amplitude-$Ro$ diagram. \label{fig:ap_general}}
\end{figure*}

The rotational modulation data used in this work are taken from \gaia\ DR2\footnote{The Gaia archive website is \href{http://gea.esac.esa.int/archive/}{http://gea.esac.esa.int/archive/}.} 
Only an abridged description of the rotational modulation data processing is given here. 
More detailed and extensive information can be found in the \gaia\ DR2 release papers, particularly in \citet{2018A&A...616A..16L,Holl_etal:2018}, and documentation \citep{2018gdr2.reptE....V,2018gdr2.reptE...7E}.
The problem of inferring stellar rotation periods from \gaia\ photometric time series, characterised by an irregular and sparse sampling, has been discussed in \cite{2012MNRAS.421.2774D}.
Considerations on period search techniques like the ones contained in, e.g., \cite{2018ApJS..236...16V} are discussed in there.

The \gaia\ DR2 rotational modulation analysis was performed by first selecting sources around the observed main sequence (MS hereafter) in the (\mg, \bmr) diagram having more than 20 field-of-view (FoV) transits.
The approximate absolute magnitude \mg\ was estimated from parallax and apparent magnitude $G$ \citep{2018A&A...616A...4E}, ignoring reddening.
Only sources with a relative parallax uncertainty better than 20\% were considered.
The observed MS was truncated at \bmr=0.6 on the blue side to limit the selection to stars of spectral type approximately later than F5. 

In order to take the photospheric active region evolution into account, the whole $G$ magnitude time series for each source was segmented using an adaptive algorithm especially tailored to the \gaia\ sampling.
This identified time series segments each covering a time interval not longer than 120\,d and including at least 12 samplings. 
The segments were allowed to partially overlap.
The maximum time span of each segment chosen arised from an unavoidable compromise between the typical timescale of active region evolution in young stars and the characteristics of the \gaia\ sampling. 
Possible spurious periods that may results from active region evolution with timescales shorter than the segment length were filtered out as described below. 
The initial list was then further limited to sources for which at least one segment could be identified.
Outliers, including possible flares, were filtered out using a robust linear regression between $G$ and \bmr\ on each segment. 

Period search was performed in each segment and in the whole time series using the generalised Lomb--Scargle periodogram method as implemented by \citet{2009A&A...496..577Z}.
The frequency range in which the search is performed was
\begin{equation}
\left(\frac{2}{T}\,, ~3.2 \right) {\rm d}^{-1} \,,
\end{equation}
where $T$ is the time interval spanned by the segment (or by the whole time series), with a frequency step $(10 T)^{-1}$~d$^{-1}$.
The upper limit 3.2\,d$^{-1}$ was chosen in order to avoid the aliases corresponding to the 6-hour rotation period of the satellite \citep[see][for a discussion on possible aliases]{2018A&A...616A..16L}.

The period with the highest power in the periodogram was selected, and the false-alarm probability (FAP), indicating the probability that the detected period is due to just noise, was associated with it.
Since Monte Carlo simulations for estimating the FAP are computationally expensive and prohibitive given the size of the \gaia\ sample \citep{2012MNRAS.421.2774D,2015MNRAS.450.2052S}, the Baluev formulation \citep{2008MNRAS.385.1279B} has been adopted.
For DR2, valid periods were chosen as those with FAP$\le$0.05, which corresponds to a false-detection rate of about $5\%$ \citep{2015MNRAS.450.2052S}
When a significant period $P$ was detected, a sinusoidal model was fitted to the time-series segment using the Levenberg-Marquardt method \citep{Levenberg_1944,Marquardt_1963}.

Periods inferred in different segments of the same source were finally combined using a method based on the mode as statistical indicator. 
Periods that were not reproduced in at least two segments or in one segment and the whole time series, with a 20\% tolerance to take uncertainties and differential rotation into account \citep{2016A&A...591A..43D}, were disregarded.

After deriving the source period and the parameters of the sinusoidal model best fitting the data, the results were further filtered by disregarding results with a poor coverage of the folded time series ($<$40\%) and/or large gaps ($>$ 30\%) in the folded time series.
Furthermore, light curves far from being sinusoidal, like the typical eclipsing binary light curve with some fairly flat outside-eclipse shape and deep minima during eclipse, are filtered out by comparing the  modulation amplitude derived from the sinusoidal fit, $A$, with that derived from the 5th and 95th percentile of the distribution.

Neglecting possible outliers in parallax and the false-detection rate of about 5\%, the expected main contaminants are essentially limited to low-mass short-period grazing binaries or binary ellipsoidal variables whose light curves are not far from being sinusoidal. 
Their expected fraction, however, is expected to be below 1\%, making their effects on the distributions over the parameters of interest negligible \citep{2018A&A...616A..16L}.

\gaia\ DR2 contains 147\,535 rotational modulation variable candidates, and therefore constitutes  the largest catalogue of rotational periods available to date.
The spectral sensitivity is strongly dependent on the ecliptic latitude and, in general, is higher for $P \apprle$\,10\,d.
The catalogue contains, therefore, an unprecedented wealth of information on the very saturated regime, which open the possibility of a change of paradigm in our understanding of stellar rotation and photospheric inhomogeneities in the early stellar evolution.

\section{Clustering in the amplitude-period diagram}\label{sec:clustering}

\subsection{The amplitude-period density diagram} \label{sec:apd}

\begin{figure*}[ht]
\gridline{\rightfig{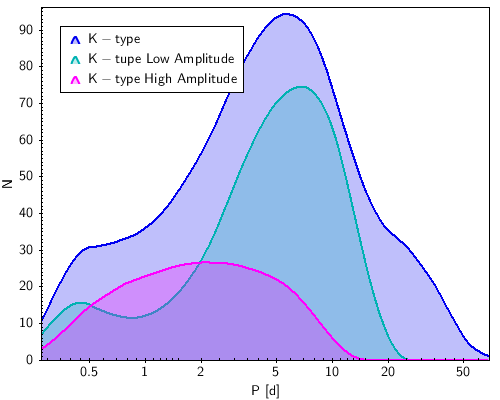}{0.455\textwidth}{(a)}
          }
\gridline{\leftfig{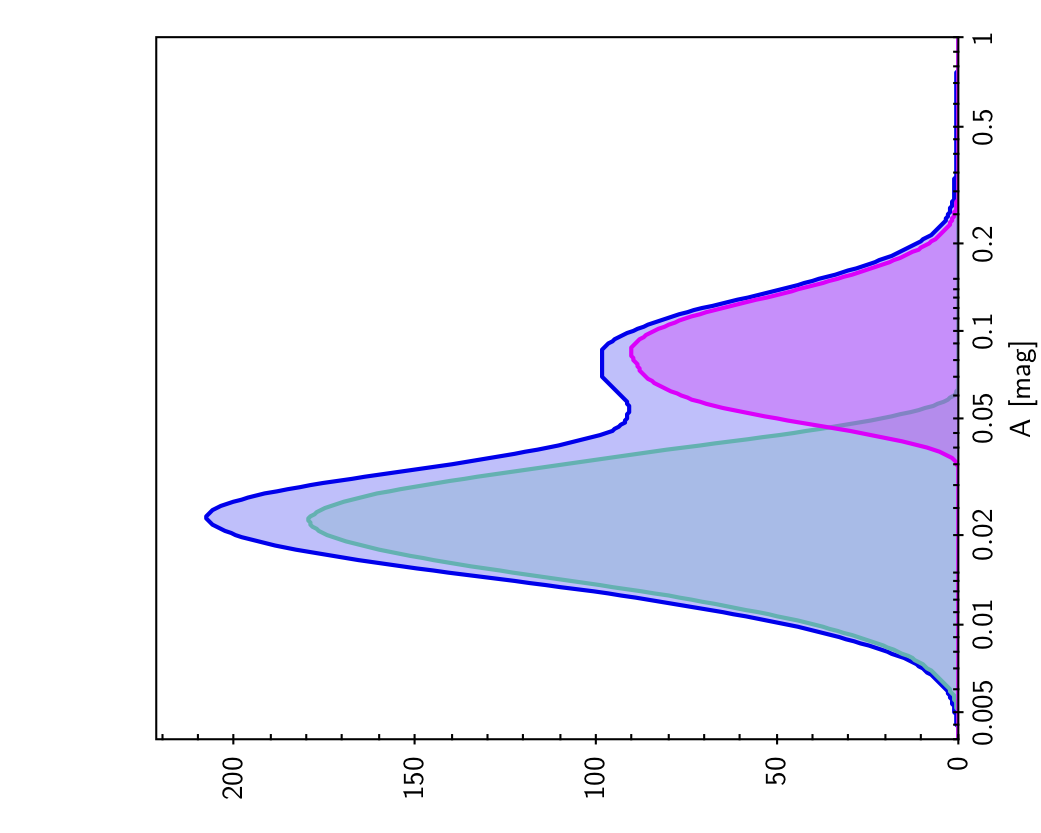}{0.495\textwidth}{(b)}
          \rightfig{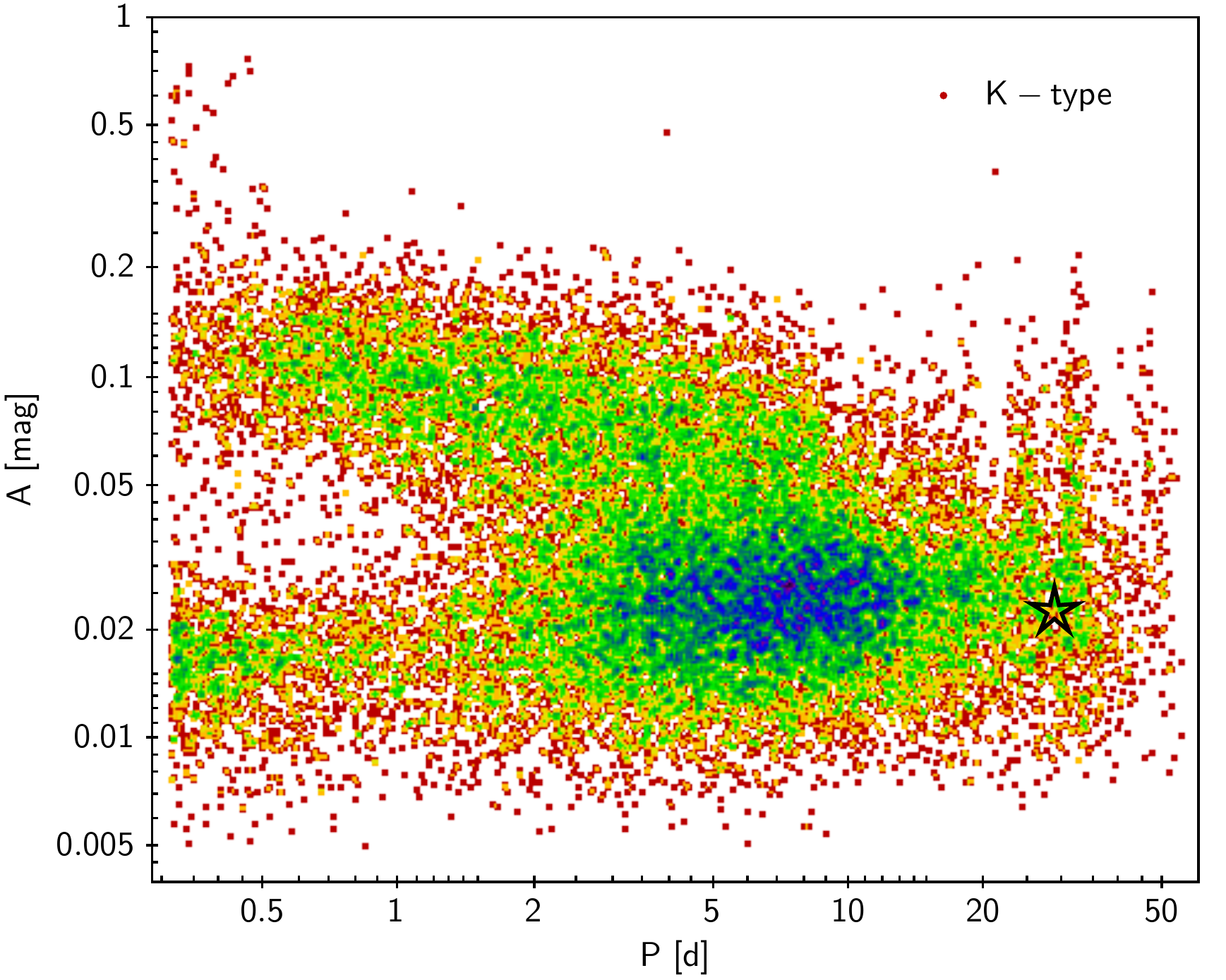}{0.48\textwidth}{(c)}
          }
\caption{Amplitude-period density diagram of K-type stars in the \gaia\ DR2 rotational modulation sample (c) with its projection in the period $P$ (a) and amplitude $A$ (b) dimensions. 
Panel (a) and (b) include the density distributions of the high- and low-amplitude groups, with the additional subdivision of the low-amplitude group into slow- and fast-rotators in panel (a).
Distributions are expressed in terms of kernel density estimate with smoothing 1.6.
\label{fig:diagrams}}
\end{figure*}

\citet{2018A&A...616A..16L} showed that the \gaia\ amplitude-period density diagram displays a major bimodality in modulation amplitude, with a boundary at $A\approx$ 0.04 -- 0.05\,mag, and a manifest gap between the low- and high-amplitude branches at $P\apprle2$\,d (see Fig.\,8 in \cite{2018A&A...616A..16L} and Figs.\,\ref{fig:ap_general} and \ref{fig:diagrams} in this paper).
Furthermore, the high-amplitude branch is more prominent at lower mass and is gradually depleted as we consider higher mass, until it almost disappear in the 1.15 - 1.25 $M_\odot$ range.
Conversely, the low-amplitude branch is very poorly populated for $P \apprle 2$\,d and $M \apprle 0.85 M_\odot$. 
This region gets increasingly populated at increasing mass \citep{2018A&A...616A..16L}.
Here we discuss more in-depth the characteristics of the multimodality in the amplitude-period density diagram presented in \citet{2018A&A...616A..16L} and infer some of the implications of these findings by comparing with prior information on open clusters, Kepler data and some individual well known stars in Sects. \ref{sec:pcd} and \ref{sec:literature}.

Fig.\,\ref{fig:ap_general} shows the amplitude-period density diagram for the whole \gaia\ DR2 rotational modulation variable sample.
In this diagram we identify three major clusterings: high-amplitude rotators (HAR; $A \apprge 0.05$\,mag), low-amplitude slow-rotators (LASR; $A \apprle 0.05$\,mag and $P \apprge 2$\,d), and low-amplitude fast-rotators (LAFR: $A \apprle 0.05$\,mag and $P \apprle 2$\,d). 
Ultra-Fast Rotators (UFR) are considered a sub-group of the LAFR with $P<0.5$\,d.
Rossby number ($Ro = P/\tau_c$) is evaluated adopting the convective turnover timescale $\tau_c$ of \citet{2013ApJ...776...87S} 

Fig.\,\ref{fig:diagrams} shows the amplitude-period density diagram of the K-type stars ($M_{\rm G} \approx$ 5.8 -- 7) in the \gaia\ DR2 rotational modulation variable sample. 
Focusing on the K stars allows a simpler, but still general, description of the features of the diagram.
All three groups can be easily identified in panel (c) of Fig.\,\ref{fig:diagrams}, and can be further illustrated with the help of the other panels.
By projecting the density diagram along the modulation amplitude dimension (panel b), the bimodality in $A$ is immediately apparent.
Dividing the entire sample into its low- and high-amplitude components makes the further segregation with respect to $P$ more easily identifiable.
With reference to panel (a) of Fig.\,\ref{fig:diagrams}, while the high-amplitude population has a broad unimodal distribution over $P$, the low-amplitude population is manifestly bimodal, with a minimum density at $P \approx 1$\,d.

The faster rotator end of the amplitude-period diagram reveals another interesting behaviour.
Stars close to break-up rotational velocity ($P<0.5$\,d) tend to cluster on the low-amplitude sequence, with a rather abrupt inversion of the relative density of the high- and low-amplitude branches at decreasing periods below $P\approx2$\,d (Fig.\,\ref{fig:diagrams}, panel (c)). 
Such a clustering gives an intuitive criterion for the definition of UFR as stars with low modulation amplitude and $P<0.5$\,d.

\begin{figure*}[!ht]
\begin{center}
\gridline{\leftfig{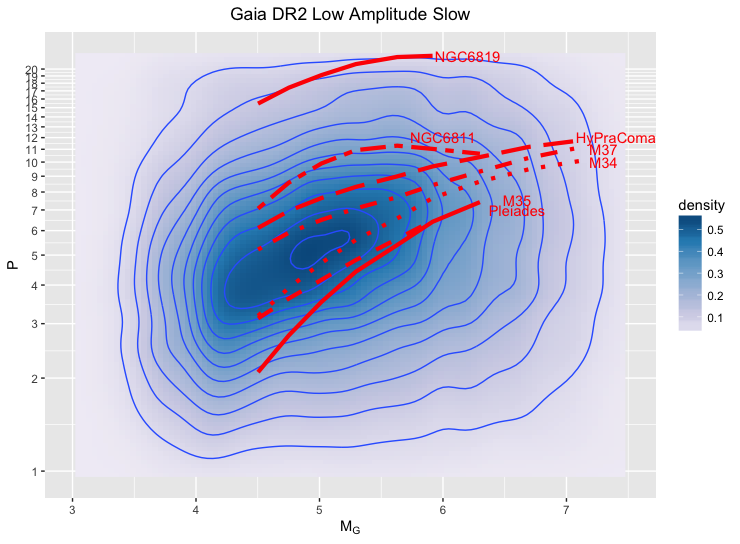}{0.48\textwidth}{(a)}
		  \rightfig{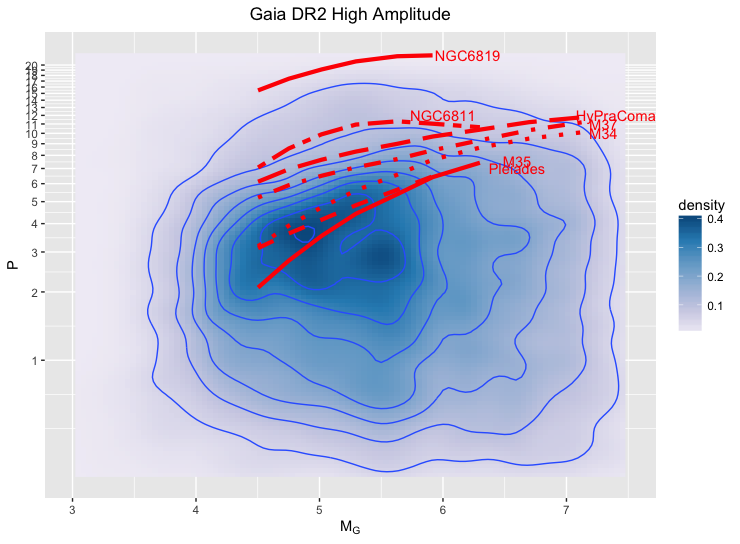}{0.48\textwidth}{(b)}
          }
\gridline{\leftfig{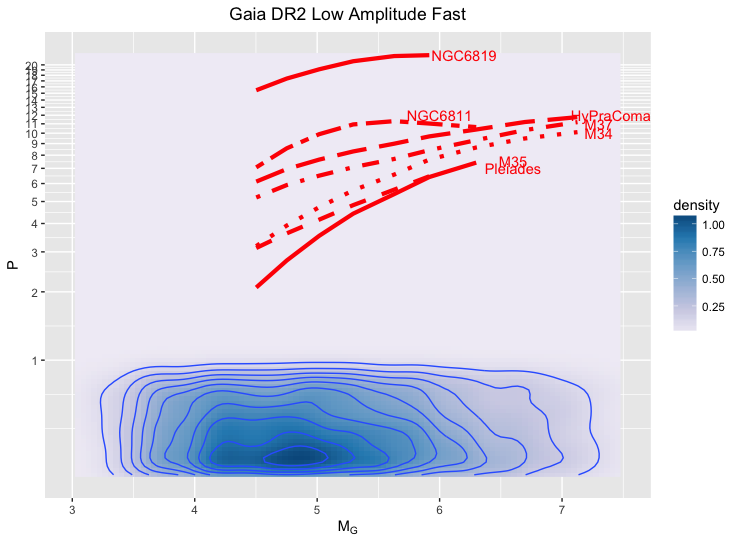}{0.48\textwidth}{(c)}
          \rightfig{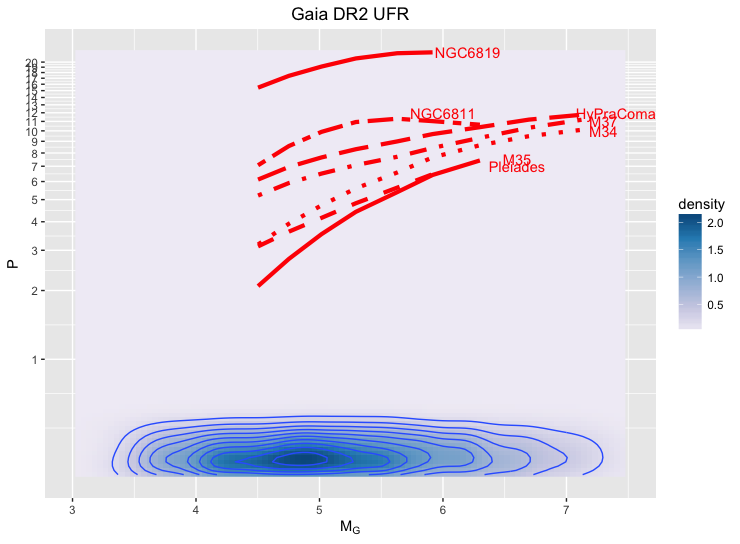}{0.48\textwidth}{(d)}
          }
\caption{\gaia\ period-magnitude density diagram (equivalent to a period-colour or period-mass diagram) for the low-amplitude slow-rotator group  compared with the open clusters slow-rotator sequence non parametric fit of \citet{2015A&A...584A..30L}. \label{fig:PM}}
\end{center}
\end{figure*}

\subsection{Relationships with the period-absolute magnitude or period-colour  diagram} \label{sec:pcd}

As already noted in \citet{2018A&A...616A..16L}, the richest group in the amplitude-period density diagram, the LASR group, corresponds to a collection of \citet{Barnes:2003} I-sequence (or slow-rotator sequence).
This is illustrated in more details in Fig.\,\ref{fig:PM}, where we show the $P$ vs unreddened absolute magnitude $M_G$ density diagram for the sources for which the extinction $A_G$ has been derived by  \citet{2018arXiv180409374A}.
The $P$ vs $M_G$ diagram is shown separately for the LASR, HAR, LAFR, and UFR groups for clarity.
As a comparison, the \citet{2015A&A...584A..30L} non-parametric fit to open clusters slow-rotator (or I-) sequence, with $(B-V)$ converted to $M_G$ using the photometric relationships of \citet{2018A&A...616A...4E} and \gaia\ DR2 parallaxes, is superimposed.
The \citet{2015A&A...584A..30L} non-parametric fit to the slow-rotator sequence was performed for each cluster in a set with ages in the 120 -- 2500 Myr range (the Pleiades, M35, M34, M37, Hyades, Praesepe, Coma, NGC6811, and NGC6819).
The identification of the slow-rotator sequence made use of a criterion derived from realisation that, when applied to older clusters with only a few fast or intermediate rotators, the non-parametric fit produces residuals whose distribution can be approximated with a normal distribution.
In order to trace back the sequence to the younger clusters, the width of the slow rotator sequence was set as the maximum width that produces a distribution of residuals with a Pearson normality probability of at least 30\%.
The final standard deviation was in the 0.35 -- 0.94 range \citep[see][for details]{2015A&A...584A..30L}.

The over-density pattern of the LASR group shows a striking similarity to the shape of the slow-rotator sequence.
It extends to higher mass than the \citet{2015A&A...584A..30L} fit, with a sharp $P$ decrease towards lower (brighter) $M_G$.
This is likely due to difficulties in fitting this part of the slow-rotator sequence, which plunges towards the faster regime (LAFR)  and is scarcely populated in any single open cluster.
At the opposite side, the density decrease at fainter/lower mass stars can be ascribed mostly to a selection effect, since the sample is not volume-limited and the more massive and brighter stars are seen at larger distance than the lower mass stars.
The decrease in density at increasing $P$ is also affected by the decrease in frequency sensitivity for $P\apprge$\,10\,d  \citep{2018A&A...616A..16L,2012MNRAS.421.2774D}.
The comparison between the LASR and the \citet{2015A&A...584A..30L} fit shows that most of the LASR stars observed by \gaia\ have rotational ages between 100 and 600 Myr.

Comparison between the LASR and the HAR groups shows that these tend to occupy different preferential location in the $P$ vs $M_G$ diagram.
The M35 slow-rotator sequence marks quite clearly the separation between the overdensity peaks of the LASR and the HAR groups.
The brighter, higher mass part of the Pleiades slow-rotator sequence (spectral type G and early K) overlaps with the HAR group.
This is as expected given the period bimodality (e.g. Fig.\,\ref{fig:diagrams}) which causes the HAR and LASR groups to overlap significantly in the period dimension for $P \approx$\,2--10\,d.
The Hyades, Praesepe, and Coma slow-rotator sequences ($\approx$600\,Myr) are located in a region of the diagram where the density of HAR stars is very low and decreases with increasing $P$ much faster than the LASR group.
The region is sufficiently below the $P\approx$\,10\,d limit, and therefore not particularly affected by variations in frequency sensitivity.
The slow-rotator sequence of NGC\,6811 (1\,Gyr) and NGC\,6819 (2.5\,Gyr), on the other hand, are located in regions essentially void of HAR stars.

Comparison between the LASR and the LAFR mirrors the behaviour depicted in Fig.\,\ref{fig:diagrams}, with a decrease in density of the LAFR as the period increases below 1\,d and an increase in density of the LASR as the period increases above 1\,d.
The LAFR density decreases at increasing $M_G$ (fainter stars) faster than the decrease of the LASR density, ruling out selection effects on the mass dependence of the LAFR group discussed in \citet{2018A&A...616A..16L}.

The density of the UFR group peaks at $M_G\approx$\,4.8\,mag and decreases faster than the HAR group at increasing $M_G$, mirroring the mass dependence of the UFR group discussed in \citet{2018A&A...616A..16L}.

\subsection{Comparison with external data} \label{sec:literature}

\begin{figure*}[tp!]
\gridline{\leftfig{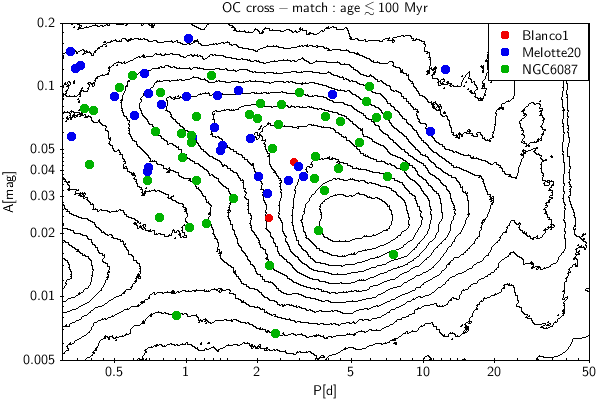}{0.46\textwidth}{(a)}
          \rightfig{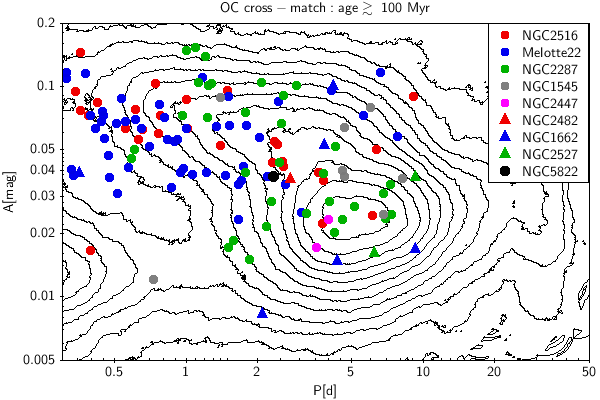}{0.46\textwidth}{(b)}}
\gridline{\leftfig{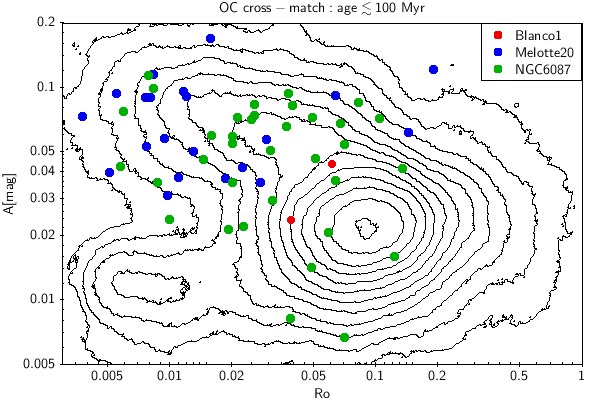}{0.46\textwidth}{(c)}
          \rightfig{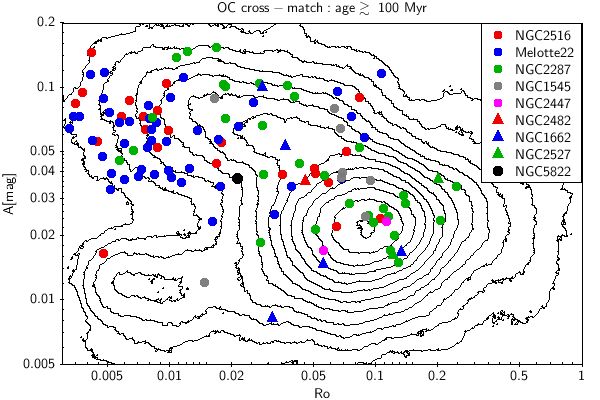}{0.46\textwidth}{(d)}}
\gridline{\leftfig{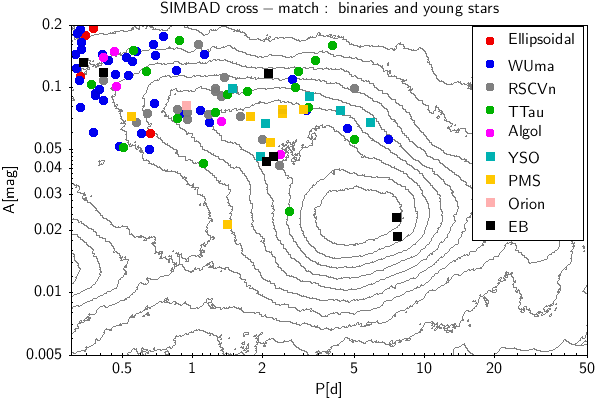}{0.46\textwidth}{(e)}
          \rightfig{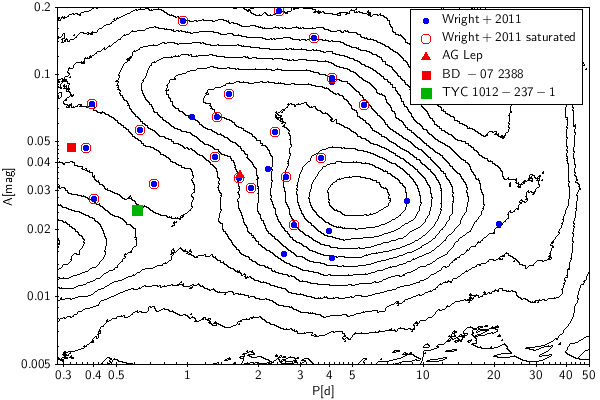}{0.46\textwidth}{(f)}}
\caption{
Panel (a): Cross-match of Gaia DR2 rotational modulation sources with members of open clusters \citep{2018A&A...618A..93C} younger than $\approx$\,100\,Myr in the amplitude-period diagram.
Panel (b): Same as panel (a) for members of open clusters older than $\approx$\,100\,Myr.
Panel (c): Same as panel (a) in the amplitude-$Ro$ diagram.
Panel (d): Same as panel (b) in the amplitude-$Ro$ diagram.
Panel (e): Cross-match with SIMBAD entries of known type.
Panel (f): Cross-match with X-ray sources \citep{2011ApJ...743...48W}. 
Panel (f) also shows the position of 
BD\,-07\,2388,
TYC\,1012-237-1,
and
AG\,Lep 
(see text fo details).
}
\label{fig:xmatch}
\end{figure*}

Hints for tracing the evolution in the $A-P$ or $A-Ro$ diagram can be provided by looking at  members of open clusters of different age.
Panels (a) and (b) in Fig.\,\ref{fig:xmatch} show the location of members of nearby open clusters \citep{2018A&A...618A..93C} with spectroscopic determination of the metallicity [Fe/H] and rotational period in \gaia\ DR2.
Clusters younger and older than $\approx$ 100\,Myr are shown in Panel(a) and (b) of Fig.\,\ref{fig:xmatch}, respectively.
Cluster ages are taken from the WEBDA database\footnote{\href{https://www.univie.ac.at/webda/}{https://www.univie.ac.at/webda/}} and used mainly to define a monotonic increase in age.

\begin{figure*}[t!]
\gridline{\leftfig{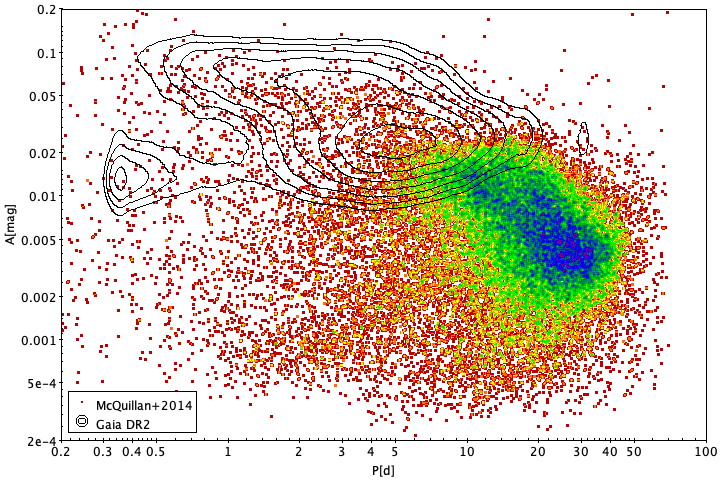}{0.46\textwidth}{(a)}
          \rightfig{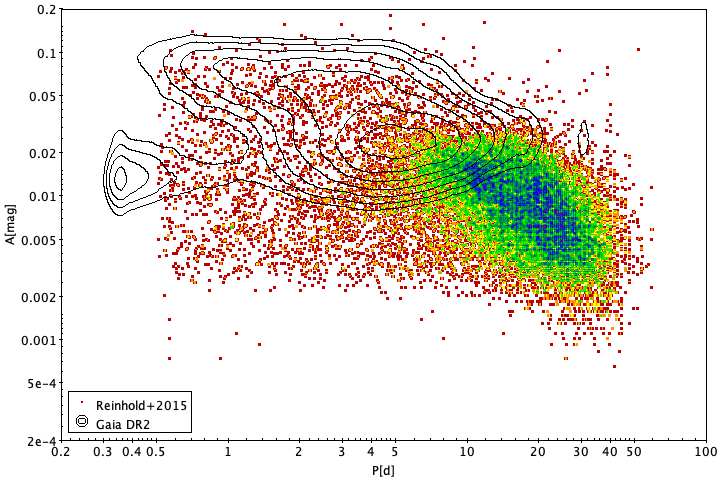}{0.46\textwidth}{(b)}}
\gridline{\leftfig{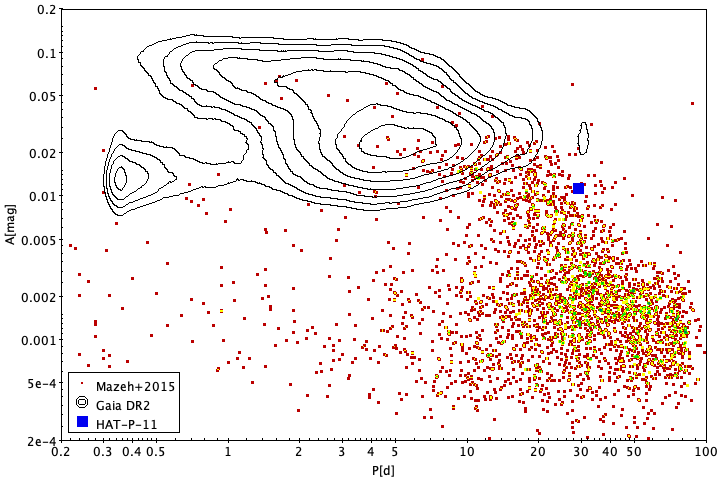}{0.46\textwidth}{(c)}
          \rightfig{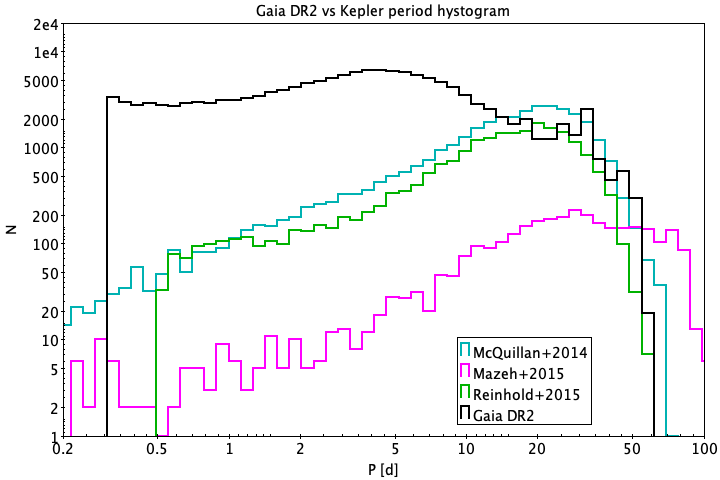}{0.46\textwidth}{(d)}}
\caption{
Comparison of Gaia DR2 rotational modulation sources (contour) with Kepler data (2D density diagram, rainbow scale). 
Panel (a): Comparison with \citealt{2014ApJS..211...24M}. 
Panel (b): Comparison with \citealt{2015A&A...583A..65R}. 
Panel (c): Comparison with \citealt{2015ApJ...801....3M}, outlining the position of HAT-P-11.
Panel (d): Period histogram (bin size 0.6\,d) for Gaia DR2 rotational modulation sources and Kepler.
Comparison with Kepler data shows the much larger density of points at short period in the gaia dataset, up to a factor $\approx$ 40 for the binning chosen here, which explains why the multimodality discovered by Gaia could not be easily recognised in the Kepler data.
}
\label{fig:kepler}
\end{figure*}

Members of clusters younger than $\approx$\,100\,Myr are essentially all on the HAR branch, with just a few of the more massive NGC6087 stars ($T_{\mathrm{eff}} \in (5000,6000)$\,K) on the LASR branch.
On the other hand, clusters older than $\approx$\,100\,Myr tend to populate the LASR branch, while the majority of their members are still on the HAR.
One star in NGC2516 and another in NGC1545, expected to be close to the ZAMS, are on the LAFR branch, the one in NGC2516 (Gaia DR2 5290815082024737664, $T_{\mathrm{eff}} = 4869$\,K) well inside the UFR region.

This picture is confirmed by the $A-Ro$ diagrams (panels (c) and (d) of Fig.\,\ref{fig:xmatch}). 
Here, the clustering on the LASR branch and the position in between the UFR and the LASR branch of Gaia DR2 270740723866954752 ($T_{\mathrm{eff}}=5362$\,K) in NGC1545 ($\approx 300$\,Myr) are even clearer.

Panel (e) in Fig.\,\ref{fig:xmatch} shows the cross-match with different types of variable as classified in the SIMBAD database\footnote{\href{http://simbad.u-strasbg.fr/simbad/}{http://simbad.u-strasbg.fr/simbad/}}.
Young objects, like those classified as T\,Tau, YSO, PMS, and Orion variable, have amplitudes $\gtrsim 0.04$\,mag, and therefore well on the HAR branch, apart one PMS and one T\,Tau.
Ellipsoidal, W\,Uma, RS\,CVn, and Algol are all located at large amplitudes.
A few eclipsing binaries are found both on the HAR and on the LASR branches.
Panel (f) in Fig.\,\ref{fig:xmatch} shows the cross-match with X-ray data from \citet{2011ApJ...743...48W} as well as the position of a few stars for illustrative purposes as discussed below.

Fig.\,\ref{fig:kepler} shows a comparison with Kepler data \citep{2014ApJS..211...24M,2015ApJ...801....3M,2015A&A...583A..65R}.
The Kepler passband is sufficiently similar to the \gaia\ G-passband to allow us a direct comparison of the modulation amplitude between the two dataset.
For such a comparison, the Kepler relative flux variation have been converted to magnitudes.

Thanks to its high-cadence sampling, Kepler data extends to much longer periods and much smaller amplitudes than the \gaia\ DR2 sample. 
On the other hand, only the \gaia\ data is so rich in the saturation regime to allow unveiling the structures in the $A-P$ diagram discussed in this work.
Apart from the much higher number of stars in the Gaia sample, the reason why the multi-modality discussed in this paper could not be seen in the Kepler data is mainly due to the different concentration of data at short and long periods, the limit at short period imposed in the \cite{2015A&A...583A..65R} dataset, the focus to the unsaturated regime in the \cite{2015ApJ...801....3M} dataset, and the limited number of stars in the UFR regime in \cite{2014ApJS..211...24M}.
The two datasets can therefore be considered complementary and cover most of the rotational evolution from the PMS to the late MS.
The most striking feature is that the over-density in the Kepler data merges with the \gaia\ LASR branch, which suggests that this latter can be identified as the tip of the unsaturated regime.
Although the scatter is very large, the well known tendency of a monotonic decrease of $A$ with increasing $P$ is evident starting from the \gaia\ LASR branch.

There is a general trend of decreasing amplitude with increasing period in the HAR branch too, with a much shallower slope than in the unsaturated regime. 
The $A$ monotonic decrease in the HAR is even more evident with respect to increasing $Ro$.
Amplitude in the LAFR branch is essentially independent from period.
Such a behaviour not only mimics the X-rays / bolometric luminosity ratio ($L_X/L_{\mathrm{bol}}$) vs period relationship \citep{2011ApJ...743...48W}, but also the  cross-match with X-ray sources presented in panel (f) of Fig.\,\ref{fig:xmatch} shows that most of the X-ray saturated sources have $(A,P)$ or $(A,Ro)$ on the HAR branch and a couple on the LAFR branch.
From a purely observational point of view, therefore, the HAR and LAFR branches can be considered as substructures of the saturated regime in rotational modulation with a close correspondence to the saturation regime seen in X-rays.

The peculiar behaviour of the very active MS star HAT-P-11 \citep{2017ApJ...846...99M,2017ApJ...848...58M} does not contradict this picture. This star is not included in \gaia\ DR2, but its amplitude and rotational period can be inferred from the Kepler dataset.
Its position on the $(A-P)$ diagram is, in fact, consistent with being at the upper edge of the unsaturated branch at the corresponding $P=29$\,d and consistent with the data scatter (Fig.\,\ref{fig:kepler}, panel (c))

Considering now the fast-rotator part of the diagram, \gaia\ DR2 does not include well known stars like AB\,Dor, BO\,Mic (Speedy Mic), and LO\,Peg as yet. 
However, conversion to the \gaia\ photometric system places their modulation amplitude on the HAR branch. AB\,Dor has $(B-V)=0.83$, $P=0.514$\,d, and $A_V = 0.13$\,mag \citep{2010A&A...520A..15M}, which correspond to $A \approx 0.08$\,mag in $G$-band using the \citet{2018A&A...616A...4E} transformations;
BO\,Mic ($(B-V)=0.94$, $P=0.38$\,d, $A_V=0.059$\,mag, \citealt{2012AcA....62...67K}) has an estimated $A\approx0.20$\,mag;
LO\,Peg ($(B-V)=1.05$, $P=0.42$\,d, $A_V=0.10$\,mag, \citealt{2010A&A...520A..15M}) has an estimated $A\approx0.21$\,mag.

All stars in the UFR branch have periods inferred for the first time in \gaia\ DR2.
The closest stars with rotational periods from previous observations are TYC\,1012-237-1 and BD\,-07\,2388.

TYC\,1012-237-1 (1SWASP\,J180500.38+111013.9) has a SuperWASP period of $P_{\mathrm{SWASP}} = 0.6070$\,d and amplitude $A_{\mathrm{SWASP}}\approx 0.036$\,mag \citep{2007A&A...467..785N}, in good agreement with the \gaia\ DR2 $P= 0.61492$\,d and $A = 0.02447$\,mag. 
\citet{2007A&A...467..785N} identified this star as the counterpart of the ROSAT source 2RXP\,J180500.8+111021.

The only well studied star close to the UFR branch, but still with an amplitude at least 0.03\,mag above its upper edge, is BD\,-07\,2388, a member of the AB\,Dor young loose association \citep{2008hsf2.book..757T} with age $\approx$\,120\,Myr. 
Its period was determined with ASAS photometry as $P_{\mathrm{ASAS}}=0.32595$\,d with an amplitude $A_V=0.037$ \citep{2012AcA....62...67K}, to be compared with the \gaia\ DR2 $P=0.32524$\,d and $A=0.047$.
By comparing the $P_{\mathrm{ASAS}}$ value with their spectropolarimetry  analysis, \citet{2018MNRAS.474.4956F} adopted $P_{\mathrm{F18}}=0.32595\pm0.0005$. 
This star has therefore a slightly shorter period than LO\,Peg and AB\,Dor, but an amplitude estimated a factor $\approx$5 smaller.
\citet{2018MNRAS.474.4956F} found its magnetic properties similar to LO\,Peg and AB\,Dor.
Its $Ro$ value is well below the limit ($Ro \approx 0.1$) for which \citet[][and references therein]{2018MNRAS.474.4956F} found evidence for saturation of the large-scale magnetic
field strength.
\citet{2018MNRAS.474.4956F} quote a surface averaged large-scale magnetic field strength $\langle B \rangle = 195$\,G.
They also estimate the percentage of axisymmetry to be 34.7\% in the poloidal, 80.8\% in the toroidal, 81.5\% in the dipole, and  62.5\% in the total components.
In comparison, LO\,Peg percentage of axisymmetry is 38.9\% in the poloidal, 48.6\% in the toroidal, 80.4\% in the dipole, and 42.5\% in the total components, with $\langle B \rangle = 139.6$\,G.  \citep{2016MNRAS.457..580F}.
The main difference between the magnetic properties of BD\,-07\,2388 and LO\,Peg is therefore the larger axisymmetry of the toroidal component of the former, which therefore may be related to its smaller modulation amplitude.
 
The position of AG\,Lep, a ``young sun'' in the Columba young loose association, is also reported  in Fig.\,\ref{fig:xmatch}, panel (f), for illustrative purposes.
\citet{2010A&A...520A..15M} inferred $P_{\mathrm{M10}}=1.895$\,d and $A_{V}=0.05$\,mag.
Its \gaia\ DR2 values, $P=1.677$\,d and $A=0.035$\,mag, place it on the HAR branch.

In this context, Doppler Imaging (DI) or Zeeman Doppler Imaging (ZDI) of selected target as well as exoplanet transmission spectroscopy techniques taking the effect of surface heterogeneity into account \citep[e.g.][]{2018MNRAS.480.5314P}, particularly around the UFR region, would add crucial information.

The comparisons presented in this Sect. are, of course bound to be based on face values, ignoring long-term variability due to, e.g., magnetic cycles.
This cannot be avoided given the data at hand and requires further confirmation by future investigations.
For the time being, however, one can deem very unlikely that long-term variability can disrupt the general trends depicted in this work.

\section{Discussion} \label{sec:discussion}

\begin{figure}[!ht]
\begin{center}
\includegraphics[width=0.45\textwidth]{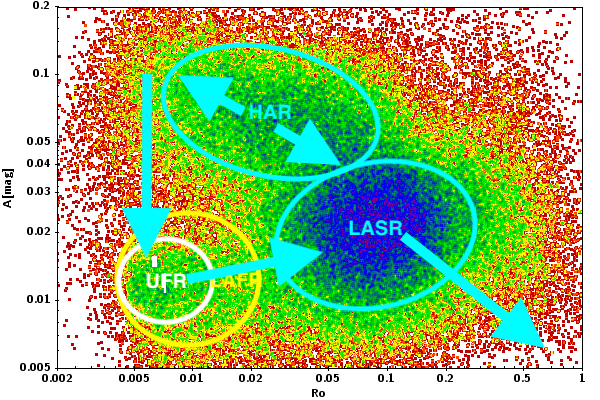}
\caption{A scenario for the low-mass stellar rotational evolution after \gaia\ DR2. \label{fig:evol}}
\end{center}
\end{figure}

Based on the new \gaia\ evidence, presented in general terms in \citet{2018A&A...616A..16L} and more in depth in Sect. \ref{sec:clustering}, and our present knowledge, in the following we propose a new scenario for the stellar magneto-rotational evolution from the T\,Tau phase to the late MS for stars that develop a core in radiative equilibrium on the MS.

\subsection{The high modulation amplitude phase}

T\,Tau stars have notoriously large photometric modulation amplitude and are naturally located in the HAR branch ($A \gtrsim 0.04$\,mag,  e.g. Fig.\,\ref{fig:xmatch}, panel (e)).
They have mean large-scale magnetic field strength $\langle B \rangle \sim 10^3$\,G with a strong mainly axisymmetric poloidal component and a much weaker toroidal component
\citep{2008MNRAS.386.1234D,2010MNRAS.409.1347D,2011MNRAS.417..472D,2013MNRAS.436..881D}.
For T\,Tau stars, $Ro\lesssim 0.1$ and $A$, $\langle B \rangle$, and $L_X/L_{\mathrm{bol}}$ are rather insensitive to $Ro$, which is what characterises the saturated regime.

In the first few million years, T\,Tau stars lose angular momentum very efficiently through the stellar-disc interaction and do not spin-up significantly despite their radius contraction.
When the angular velocity regulation of the star-disc interaction becomes inefficient, the  star spins up while it contracts approaching the ZAMS \citep[e.g.][and references therein]{2013A&A...560A..13M,2018AJ....155..196R}.

Considering stars that are sufficiently massive to develop a radiative core, 
the mean large-scale magnetic field strength $\langle B \rangle$ decreases to a few $10^2$\,G and the poloidal component becomes less axisymmetric \citep{2016MNRAS.457..580F}.
For $Ro \lesssim 0.1$, however, the magneto-rotational evolution is still very uncertain, and the $A$ and $\langle B \rangle$  scatter with rotation or $Ro$ is very large.
The \gaia\ evidence, combined with current theories \citep[e.g.][and references therein]{2013A&A...556A..36G,2015A&A...577A..98G}, suggest that in this phase the star moves to the left along the HAR branch until it stops contracting once on the ZAMS.
After the ZAMS the star spins down as it loses angular momentum via the magnetised wind and it moves to the right along the HAR branch towards increasing $Ro$ and the LASR branch (Fig.\,\ref{fig:evol}).

\subsection{From the high-amplitude to the low-amplitude slow-rotator branch}
\label{sec:har2lasr}

When a star on the HAR branch approaches the HAR lower-right boundary in the $A-P$ or $A-Ro$ diagram, Fig.\,\ref{fig:diagrams} suggests that it transits quite rapidly to the LASR branch.
Assuming that this latter is the tip of the over-density defined by the Kepler data (Fig.\,\ref{fig:xmatch}, panel (f)), $A$, $\langle B \rangle$ , and $L_X/L_{\mathrm{bol}}$  correlation with $P$ or $Ro$ becomes higher (larger slope), more regular, monotonic, although still affected by a large scatter \citep[e.g.][]{2016MNRAS.457..580F,2018MNRAS.474.4956F,2011ApJ...743...48W}.
In other words, the star enters into the unsaturated regime.
The small gap between the HAR and LASR branches indicates that the transition from the saturated to the unsaturated regime is somewhat discontinuous, as some sort of transition in surface inhomogeneities would imply.
Fig\,\ref{fig:ap_general} shows that this transition takes place with $Ro \in (0.03, 0.13)$.
A transition of this sort is rather poorly described by the \citet{1995ApJ...441..865C,1995ApJ...441..876C} saturation level or by the \citet{Krishnamurthi_etal:1997} scaling law Eq.\,(\ref{eq:Krishna_scaling}) and (\ref{eq:Krishna_threshold}).
The limitation of such type of relationships implies also that wind-braking laws based on a simple $Ro$ threshold between the saturated and unsaturated regimes, like Eq.\,(\ref{eq:Matt15}), have limited validity and bear inconsistencies.

One of such inconsistencies has been outlined by \citet{2015A&A...584A..30L} (see also Sect.\,\ref{sec:intro}), who demonstrated that the wind braking law Eq.\,(\ref{eq:LS15}) is valid for the slow-rotator sequence, without assuming a well defined $Ro$ threshold for the saturated/unsaturated regimes.
On the other hand, wind-braking laws of the Eq.\,(\ref{eq:Matt15}) type, based on a well defined $Ro$ threshold for the saturated/unsaturated regime, imply that both saturated (lower mass) and unsaturated (higher mass) stars are present on the slow-rotator regime and, viceversa, more massive unsaturated stars may have not yet settled on the slow-rotator regime.
In summary, Eqs.\,(\ref{eq:Krishna_scaling}), (\ref{eq:Krishna_threshold}), and wind-braking laws of the Eq.\,(\ref{eq:Matt15}) type are incompatible with the identification of the slow-rotator sequence with the unsaturated regime.

Another inconsistency borne by Eqs.\,(\ref{eq:Krishna_scaling}), (\ref{eq:Krishna_threshold}), and wind-braking laws of the Eq.\,(\ref{eq:Matt15}) type is their incompatibility with the $A$ bimodality with respect to $Ro$ in the range where the transition from the HAR to the LASR branches takes place (e.g., Fig.\,\ref{fig:ap_general}).
This, together with the small gap between the HAR and the LASR branches, rules out a deterministic description of the transition between the saturated and unsaturated regimes and supports the idea of a stochastic transition of the type proposed by \citet{2014ApJ...789..101B}.

\subsection{From the high-amplitude to the ultrafast rotator branch}

The \gaia\ evidence also indicates an alternative magneto-rotational evolution track
for those stars that get close to breakup velocity when contracting towards the ZAMS.
The presence of ultrafast rotators in open clusters with members close to the ZAMS has long been established \citep[e.g.][]{1985ApJ...289..247S,1993ApJ...409..624S}.
The very existence of the UFR branch in the $A-P$ diagram and the density inversion between the HAR and the UFR branches at $P\lesssim0.5$\,d with a clear gap in between the two branches in, e.g., Fig.\,\ref{fig:diagrams} strongly suggest that in this case the star may undergo a very rapid transition from the HAR to the UFR, with a sudden drop of about a factor 10 in modulation amplitude.
The cross-match with young open clusters presented in Fig.\,\ref{fig:xmatch} panels (a)-(d) provides further support to the existence of such a transition.
Contrasting the position of members of clusters younger than 100\,Myr with that of the members of older clusters, there is an increase of the relative number of stars at the faster end of the diagram, as well as an increase in the relative number of stars in the LASR branch, and the first appearance of a couple of UFR.

We have no information as yet on what could cause such a sudden modulation amplitude drop and what is the contrast and distribution of surface inhomogeneities that leads to such a dramatic decrease in modulation amplitude.
The closest regime fairly well studied to date is that of very fast rotators like BO\,Mic, LO\,Peg, AB\,Dor, and BD\,-07\,2388, that have, nevertheless, sufficiently high modulation  amplitude to belong to the HAR branch.
Although there is no detailed study of stars that can be placed on the UFR branch presented here, a hint may be provided by contrasting the magnetic properties of LO\,Peg and AB\,Dor with those of BD\,-07\,2388, which has a modulation amplitude a factor $\approx 5$ smaller than the other two, placing it in between the HAR and the UFR branch.
As discussed in Sect.\,\ref{sec:literature}, the main difference in the magnetic properties of BD\,-07\,2388 with respect to LO\,Peg and AB\,Dor is the higher degree of axisymmetry of the toroidal component of the magnetic field.
Assuming that BD\,-07\,2388 is representative of stars transiting from the HAR to the UFR branch, this suggests that the small modulation amplitude of UFR stars may be due to a high degree of axisymmetry of the magnetic fields, although this suggestion obviously awaits confirmation by further detailed studies.

\subsection{The low-amplitude fast-rotator branch}

After the transition from the HAR to the UFR branch, if the magnetic configuration had a one-to-one correspondence with the rotation velocity, the star would return very rapidly to the HAR branch as it spins down below the $P\approx 0.5$\,d limit.
Then, from there, the star would make its path towards the LASR as the other HAR stars.
However, the 2D density distribution suggests an alternative scenario.
The UFR region is connected with the LASR sequence by a rather low density bridge, while it seems almost completely disconnected from the HAR branch.
This suggests that, while there is no alternative for a ZAMS star that becomes UFR but to jump abruptly from the HAR branch to the UFR over-density region, there is a non-negligible probability that the UFR makes then its way directly to the LASR sequence on the low-amplitude branch without returning to the HAR branch first.

The existence of this alternative evolutionary path strengthen the idea that there is no simple one-to-one relationship between magnetic configuration and rotation on the early main-sequence, which has an analogy with the high-amplitude -- low-amplitude bimodality at $P > 2$\,d.
Furthermore, this would imply also a dependence on the stellar rotation history of the surface magnetic field configuration.

\subsection{Evolution on the low-amplitude slow-rotator branch}

Once the star enters in the LASR branch, $A$ (Fig.\,\ref{fig:xmatch}), $\langle B \rangle$ \citep{2016MNRAS.457..580F,2018MNRAS.474.4956F}, and $L_X/L_{\mathrm{bol}}$ \citep{2011ApJ...743...48W} correlation with $P$ or $Ro$ becomes higher (larger slope), more regular, monotonic, although still affected by a large scatter.
In other words, the star enters into the unsaturated regime.
On the $A-P$ diagram, Fig.\,\ref{fig:xmatch} panel (f) suggests that on the LASR stars evolve  down and to the right along the Kepler over-density.
Comparison with the \citet{2015A&A...584A..30L} fit to the slow-rotator sequence (Fig.\,\ref{fig:PM}) indicates that here the wind-braking law Eq.\,(\ref{eq:LS15}) reproduces satisfactorily the angular velocity evolution and its dependence on mass, provided core-envelope angular momentum coupling is taken into account.

\subsection{Stars with deep or shallow convective zone on the main sequence}

The very low density of $M \sim 0.7 M_{\odot}$ stars in the LAFR branch compared to the HAR branch shown in the $A-P$ diagram of Fig.\,8 of \citet{2018A&A...616A..16L} or the faster LAFR density decrease than HAR at decreasing mass in Fig.\,\ref{fig:PM} (increasing $M_G$) suggest that stars with deep convective zone on the MS have a very low probability to transit from the HAR to the UFR branch.
Once these stars spin-down on the HAR approaching the HAR lower-right boundary in the $A-P$ or $A-Ro$ diagram, they transit quite rapidly to the LASR branch like other stars, as suggested by the gap between the HAR and LASR branches.
 
Conversely, the low density of $M \sim 1.1 M_{\odot}$ stars in the HAR branch of Fig.\,8 of \citet{2018A&A...616A..16L} or the comparison of the LAFR with the HAR density at increasing mass (decreasing $M_G$) in Fig.\,\ref{fig:PM} suggest that stars having rather shallow convection zone on the MS ($M \sim 1.1 M_{\odot}$) have a high probability of switching to the UFR (or to the LAFR) branch as they contract towards the ZAMS.
On the LAFR branch these stars evolve towards increasing periods and, once on the LASR, they evolve as the other unsaturated stars. 

No sufficient statistical information is available as yet on stars that remain fully convective on the MS to infer their possible magneto-rotational evolution in the context of the \gaia\ evidence.
Possible connections with the work of \cite{2017ApJ...834...85N}, who observed a threshold in the mass-period plane that separates active and inactive M dwarfs, is therefore still to be explored.

\section{Conclusions}\label{sec:conclusions}

The data of rotational modulation variables contained in \gaia\ DR2 reveal, for the first time, signatures of different regimes of their surface inhomogeneities in the first 600\,Myr of their evolution.
These signatures are evident in the amplitude-period density diagram, where we identify three major clusterings: high-amplitude rotators (HAR); low-amplitude slow-rotators (LASR); low-amplitude fast-rotators (LAFR).
The \gaia\ very detailed amplitude-period density diagram permits also an intuitive and rigorous definition of ultra-fast rotators (UFR) as a sub-group with $P<0.5$\,d of the LAFR group. 

The manifest segregation of these three groups in the amplitude-period density diagram, together with the prior knowledge that all stars end up in the low-amplitude slow-rotator group as they age, hints at rather rapid transitions from one regime to another.
In other words, stars change their appearance quite rapidly at some stage of their early evolution and some of them even more than once.

The transition from the high-amplitude group to the low-amplitude slow-rotator group has a correspondence to the \citet{Barnes:2003} transition from the C-sequence to the I-sequence through a gap phase or the equivalent transition in the metastable dynamo models of \citet{2014ApJ...789..101B}.
What is entirely new is the possible evolutionary path leading to the UFR conditions and then, from such conditions, to the low-amplitude slow-rotator regime.
The \gaia\ amplitude-period diagram suggests that stars passing through the UFR regime undergo first a very rapid transition from the high-amplitude branch to the low-amplitude branch when they contract and spin-up to almost break-up rotational velocity ($P<0.5$\,d).
Then, as the star spins down, it remains along the low-amplitude branch and, when $P>0.5$\,d, it passes quite rapidly from the low-amplitude fast-rotator group to the final low-amplitude slow-rotator group.

The \gaia\ evidence, compared to the Kepler data and cross-matched with well studied stars, supports the identification of the \gaia\ LASR branch with the tip of the unsaturated regime and of this latter with the slow-rotator sequence in \citet{2018A&A...616A..16L} or, equivalently, with the \citet{Barnes:2003} I-sequence.
In the unsaturated regime, the \citet{2015A&A...584A..30L} wind-braking law (Eq.\,\ref{eq:LS15}) reproduces satisfactorily the angular velocity evolution and its dependence on mass, provided core-envelope angular momentum coupling is taken into account.
A deterministic threshold between the saturated and unsaturated regime \citep{1995ApJ...441..865C,1995ApJ...441..876C,Krishnamurthi_etal:1997} is inconsistent with the clustering observed in the \gaia\ $A-P$ or $A-Ro$ diagram and with the identification of the slow-rotator sequence with the unsaturated regime.
Conversely, a stochastic transitions from fast- to slow-regimes, as that proposed by \citet{2014ApJ...789..101B}, is more consistent with the \gaia\ evidence.

Finally, the amplitude multimodality can be exploited to remove the $P$ degeneracy in field stars, for which a period-colour diagram cannot be used to identify the \citet{Barnes:2003} I-sequence to which gyrochronology relationships can be applied.
In fact, stars with low-amplitude modulation ($A \apprle 0.05$\,mag) and $P\apprge 2$\,d are expected to belong to the unsaturated population and therefore their age can be estimated from their rotational period, provided a useful proxy of the star mass and an estimate of the metallicity are available.

\acknowledgments
This work makes use of results from the European Space Agency (ESA) space mission Gaia. 
Gaia data are being processed by the Gaia Data Processing and Analysis Consortium (DPAC). 
Funding for the DPAC is provided by national institutions, in particular the institutions participating in the Gaia MultiLateral Agreement (MLA). 
This work was supported by the Italian funding agencies Agenzia Spaziale Italiana (ASI) through grants I/037/08/0, I/058/10/0, 2014-025-R.0, and 2014- 025-R.1.2015 to INAF 
(PI M.G. Lattanzi). 
The Gaia mission website is \href{https://www.cosmos.esa.int/gaia}{https://www.cosmos.esa.int/gaia}. 
The Gaia Archive website is \href{http://gea.esac.esa.int/archive/}{http://gea.esac.esa.int/archive/}.

%

\vspace{5mm}
\facility{ESA-Gaia}


\software{R (\href{https://www.r-project.org}{https://www.r-project.org}), Topcat \citep[\href{http://www.starlink.ac.uk/topcat/}{http://www.starlink.ac.uk/topcat/},][]{2005ASPC..347...29T}}

\bibliographystyle{aasjournal}
\bibliography{Proteus}

%
%



\end{document}